\documentclass[aps,prl,nofootinbib,showpacs,twocolumn,superscriptaddress]{revtex4-1}

\usepackage[english]{babel}
\usepackage[utf8]{inputenc}
\usepackage{amsmath}
\usepackage{amssymb}
\usepackage[caption=false]{subfig}
\usepackage{amssymb}
\usepackage{epsfig}
\usepackage{graphicx}
\usepackage{amsmath}
\usepackage{array,color}
\usepackage{natbib}

\usepackage[usenames,dvipsnames]{xcolor}
\definecolor{forestgreen}{rgb}{0.11,0.54,0.15}
\definecolor{purple}{rgb}{0.62,0.10,0.96}
\definecolor{dockerblue}{rgb}{0.11,0.56,0.98}
\definecolor{freeblue}{rgb}{0.25,0.41,0.88}

\usepackage[pdftex,plainpages=false,colorlinks=true,linkcolor=Red, citecolor=blue, urlcolor=blue]{hyperref}

\begin{document}

\title{Evolution of the cyclotron mass with doping in La$_{2-x}$Sr$_x$CuO$_4$}

\author{A. Legros}
\affiliation{Department of Physics and Astronomy, The Johns Hopkins University, Baltimore, MD 21218 USA.}

\author{K.W. Post}
\affiliation{National High Magnetic Field Laboratory, Los Alamos National Laboratory, Los Alamos, NM 87545}

\author{Prashant Chauhan}
\affiliation{Department of Physics and Astronomy, The Johns Hopkins University, Baltimore, MD 21218 USA.}

\author{ D. G. Rickel }
\affiliation{National High Magnetic Field Laboratory, Los Alamos National Laboratory, Los Alamos, NM 87545}

\author{Xi He}
\affiliation{Brookhaven National Laboratory, Upton, NY 11973, USA.}
\affiliation{Department of Applied Physics, Yale University, New Haven, Connecticut 06520, USA.}

\author{Xiaotao Xu}
\affiliation{Brookhaven National Laboratory, Upton, NY 11973, USA.}
\affiliation{Department of Physics, The University of Texas at Dallas, Richardson TX 75080-3021, USA}

\author{Xiaoyan Shi}
\affiliation{Department of Physics, The University of Texas at Dallas, Richardson TX 75080-3021, USA}

\author{Ivan Bo{\v{z}}ovi{\'{c}} }
\affiliation{Brookhaven National Laboratory, Upton, NY 11973, USA.}
\affiliation{Department of Applied Physics, Yale University, New Haven, Connecticut 06520, USA.}

\author{S.A. Crooker}
\email{crooker@lanl.gov}
\affiliation{National High Magnetic Field Laboratory, Los Alamos National Laboratory, Los Alamos, NM 87545}

\author{N.~P.~Armitage}
\email{npa@jhu.edu}
\affiliation{Department of Physics and Astronomy, The Johns Hopkins University, Baltimore, MD 21218 USA.}
\affiliation{Canadian Institute for Advanced Research, Toronto, Ontario M5G 1Z8, Canada.}

\date{\today}
\begin{abstract}
The recent observation of cyclotron resonance in optimally-doped La$_{2-x}$Sr$_x$CuO$_4$ using time-domain THz spectroscopy in high magnetic field has given new possibilities for the study of cuprate superconductors. One can measure the cyclotron mass in the more disordered cuprates possessing short scattering times therefore expanding the study to materials and dopings in which quantum oscillations have not been observed.  Here we present the measurement of the carrier mass of the hole doped cuprate La$_{2-x}$Sr$_x$CuO$_4$ across a range of dopings spanning from the slightly underdoped ($p=0.13$) to highly overdoped ($p=0.26$), near the termination of the superconducting dome.  These results reveal a systematic increase of $m_c$ with doping, up to values greater than thirteen times the bare electron mass.   This is in contrast with the masses extracted from the heat capacity, which show a peak near the pseudogap critical point $p^*$ and/or Lifshitz transition.  The cyclotron frequency is linear in field up to 31 T for all dopings giving no evidence for field-induced Fermi surface reconstructions.  The cyclotron mass is found to be positive for all dopings, but with a magnitude systematically below the heat capacity mass for under and optimally doped samples, while exceeding it for overdoped samples.  Among other aspects, these results are surprising as photoemission reveals a Lifshitz transition in the middle of our doping range and the sign of the cyclotron mass determined from a finite frequency resonance is -- in conventional theories -- a {\it topological quantity} only sensitive to whether or not the Fermi surface is closed around holes or electrons.  We see no sign of a divergence of the mass near $p^*$ nor near the Lifshitz transition, showing that any singularity -- if it exists -- is not strong enough to affect the cyclotron mass.
\end{abstract}
\maketitle

\section{I. Introduction}


The microscopic nature of the pseudogap phase, its critical point $p^*$, and the dominant interactions in the hole-doped cuprates~\cite{keimer_quantum_2015} are still major topics of inquiry. In the vicinity of the pseudogap quantum critical point many intriguing properties are found, such as a \textit{T}-linear resistivity down to \textit{T}~$\rightarrow$~0~\cite{daou_linear_2009,cooper_anomalous_2009}, a large increase of the effective mass~\cite{michon_thermodynamic_2019}, a logarithmic divergence as a function of temperature of the specific heat~\cite{michon_thermodynamic_2019} and Seebeck coefficient~\cite{daou_thermopower_2009,lizaire_transport_2021-1} along with a crossover in the carrier density from $p$ to  $1+ p $ with increasing doping~\cite{badoux_change_2016,collignon_fermi-surface_2017}. Some of these signatures are reminiscent of quantum criticality, but the situation in cuprates seems to be more complex than in quantum critical materials such as heavy fermions, organics or iron pnictides as some of these properties have been observed over a wide range of dopings.  Clarifying the role of the critical doping $p^*$ and the dominant interactions is crucial for establishing a model of the cuprates.

The recent observation of cyclotron resonance (CR) in the  circularly-polarized complex conductivity of an optimally-doped La$_{2-x}$Sr$_x$CuO$_4$ (LSCO) film using time-domain THz spectroscopy in pulsed magnetic field~\cite{post2021observation} led to the measurement of the cyclotron mass in a superconducting cuprate.  Cyclotron motion is the driven resonant motion of mobile charge carriers subject to a Lorentz force.  In a conventional metallic system it occurs at frequency $\omega_c = eB/m_c$, where $m_c$ is the cyclotron mass, which is a quantity generally close to, but not precisely the same as the mass inferred from other techniques.  CR has the advantage over techniques like quantum oscillations in that it can be measured even when the cyclotron frequency $\omega_c$ is much less than the carrier scattering rate $\Gamma = 1/\tau$.  The ability of the CR technique to measure masses at relatively high temperatures and modest magnetic fields (up to 50K with fields up to 31T) as compared to quantum oscillation experiments demonstrates that CR is an alternative for the direct extraction of crucial quantities such as the scattering rate and mass of charge carriers in correlated systems despite small $\tau$.  With a fit of the \textit{ac} conductivity based on the Drude model, one can obtain these parameters at a specific field and temperature without relying on the Lifshitz-Kosevich formalism.

CR is an established technique that was first developed to study semiconductors in the 1950s~\cite{dresselhaus1955cyclotron}, and proved important for probing 2D electron gases, fractional quantum Hall systems, topological semi-metals and surface states of topological insulators~\cite{jiang2007infrared,prashant2021,crassee2011giant, wu2015high,cheng2019magnetoterahertz}.  However, the technique has only been applied to a small number of correlated systems through the use of microwaves rather than THz~\cite{singleton1992far, kimata2011cyclotron,tonegawa2012cyclotron,hill2000cyclotron}.  In this regard, the higher THz-range frequencies and larger (pulsed) magnetic fields described in Ref.~\cite{post2021observation} opened the door to the CR study of high-temperature superconductors.

In addition to extracting $\omega_c$ and $ \Gamma$, such studies enable the comparison of effective masses extracted from different techniques.  In the non-interacting limit the cyclotron mass $m_c$ is simply expressed~\cite{ashcroft1976introduction} as $m_c = \frac{\hbar^2}{2 \pi} \frac{\partial A}{\partial E}$, where $A$ is the cross-sectional area of the Fermi surface in the plane normal to the applied field.  Electron-electron correlations can influence the effective masses measured by cyclotron resonance differently~\cite{kanki1997theory} as compared to other measures of the thermodynamic and quasiparticle masses (for example quantum oscillations, specific heat measurements, and angle-resolved photoemission).  Kohn's famous proof~\cite{kohn1961cyclotron} that in a Galilean invariant system electron-electron interactions do not modify the cyclotron mass is not relevant to the cuprates, which have nearly half-filled bands and feel the lattice strongly.  Moreover, finite disorder and non-parabolicity can cause electron-electron interactions to manifest in CR experiments~\cite{macdonald1989cyclotron}.  However Kohn's result does establish that CR can be sensitive to interactions in a different fashion than other probes.  Therefore comparing the cyclotron mass to masses extracted with different techniques may bring new insight into the dominant interactions in these compounds.

Here we expand our study by measuring CR on films that range from slightly underdoped ($p=0.13$) to highly overdoped ($p=0.26$), close to the end of the superconducting dome, therefore spanning the regions above and below $p^*$ (located approximately at $p^*$ = 0.19~\cite{cooper_anomalous_2009}). 
Until recently, regions of charge density waves (CDW) and spin density waves (SDW) were thought to be constrained to the underdoped region of the phase diagram, but X-ray diffraction studies on LSCO have revealed the presence of CDW correlations at $x$ = 0.16~\cite{wen_observation_2019} and even more recently up to at least $x$ = 0.21~\cite{miao2021charge}, while NMR and ultrasound measurements in high magnetic fields unveiled quasi-static magnetism persisting up to $p^*$ in strong field~\cite{frachet2020hidden}. CR should be sensitive quite generically to various density wave states that reconstruct the Fermi surface~\cite{chakravarty2001hidden}.  Looking at samples that span the slightly underdoped with $T_c$ = 39K to overdoped with $T_c$ = 5K, we observe a hole-type cyclotron resonance effect in all samples with an increasing trend of the cyclotron mass with doping.   We find a quantitative disagreement with the carrier mass extracted from heat capacity studies.  The cyclotron mass is lower than that extracted with heat capacity for $p<0.25$, but is larger for $p>0.25$.   Moreover, the hole-type cyclotron resonance found across the measured dopings is unexpected considering the hole- to electron-like Lifshitz transition at $x\approx 0.205$ ~\cite{yoshida_systematic_2006,Zhong2022}.   Such experiments may be useful in determining the dominant interactions in the cuprates.


\section{II. Methods}

The samples presented in this work are La$_{2-x}$Sr$_x$CuO$_4$ thin films of various compositions and   $T_c$, listed in Table~\ref{samples}. Nine samples were grown by atomic layer-by-layer molecular beam epitaxy on a 1 mm thick (001)-oriented LaSrAlO$_4$ substrate and characterized by mutual inductance and Reflection High-Energy Electron Diffraction (RHEED). Details of the thin film growth can be found in Ref.~\cite{bozovic_dependence_2016}.  Data on the $x$ = 0.16 sample were reported in Ref.~\cite{post2021observation}. For LSCO bulk crystals, one usually uses the nominal doping (Sr concentration) as the hole doping to locate the samples in the phase diagram.  But in the case of thin films, depending on how oxidizing the sample growth environment, the presence of excess oxygen leads to variations in $T_c$ between samples of the same nominal Sr composition. Therefore, to estimate the hole doping of our samples and  their positions in the phase diagram, we use the $T_c$ rather than the nominal Sr doping $x$ to estimate the carrier density. The absolute values of $p$ are determined based on the conjectured relation between $T_c$ and $p$ proposed in Ref.~\cite{presland_general_1991}, using $T_{c,max}$ = 41K.  A recent ARPES study has shown that this relation predicts the inferred doping for both films and crystals from the Luttinger volume to within $p= \pm 0.015$~\cite{Zhong2022}.

\begin{table}[t]
\begin{center}
\begin{tabular}{l|ccc|ccccc}
\hline
Sample name & \textit{x} & $T_c$ & \textit{p} & $\#$ of layers & Thickness $t$ \\
 &  & (K) & &  & (nm) \\
\hline
\hline
LSCO UD39K & 0.14 & 39 & 0.131 & 20 & 13  \\
LSCO OP41K & 0.16 & 41 & 0.16 & 80 & 53  \\
LSCO OD36K & 0.19 & 36 & 0.202 & 20 & 13  \\
LSCO OD35K & 0.175 & 35 & 0.205 & 20 & 13  \\
LSCO OD32K & 0.25 & 32.5 & 0.212 & 60 & 40 \\
LSCO OD26K & 0.25 & 26  & 0.228 & 20 & 13 \\
LSCO OD17K & 0.28 & 17.5 & 0.244 & 20 & 13 \\
LSCO OD13K & 0.27 & 13.5 & 0.251 & 20 & 13 \\
LSCO NSC & 0.32 & 5 & 0.263 & 20 & 13 &   \\
\hline
\end{tabular}
\end{center}
\caption{La$_{2-x}$Sr$_x$CuO$_4$ thin films used in this work.   $x$ is the nominal doping level and $p$ is the doping level inferred from $T_c$ and the relation in Ref.~\cite{presland_general_1991}.  The most overdoped sample (x = 0.32) is labeled NSC due to a lack of superconductivity signature in its THz spectrum down to the lowest temperatures we could measure. High-field data on LSCO OP41K have already been published in Ref. \cite{post2021observation}.
}
\label{samples}
\end{table}

The conditions to observe CR in cuprate thin films have been challenging to implement because of their relatively large masses (as compared to classic semiconductor work) and large scattering rates.   In order to resolve a frequency shift the cyclotron resonance $\omega_c = eB/m_c$ must be an appreciable fraction of the scattering rate $\Gamma$.   Moreover measurement frequencies $\omega$ of the order or slightly greater than $\omega_c$ are required.   With scattering rates in the THz range, and masses greater than the free electron mass, magnetic fields in 10s of Tesla and the THz spectral range must be used.  Finally, as the cyclotron frequency shifts are -- despite the field -- still small, it is superior to analyze the complex conductivity in the $r$ and $l$ circular bases (for right-handed and left-handed circularly polarized light respectively).  Measurements of the full complex optical conductivity matrix (in $r/l$ or $x/y$ bases) in magnetic field are only possible with a phase sensitive technique like time-domain THz experiments\footnote{Measurements of the magnitude of the reflectivity in the $x/y$ bases as performed in Fourier Transform InfraRed (FTIR) experiments do not allow the quantification of small cyclotron frequency shifts by themselves.  Note that the Kramers-Kronig relations usually used to determine phase information in FTIR do not work if the eigenbasis for reflection or transmission is circular (as it will be for LSCO in $c$-axis magnetic field), but one measures in the $x/y$ basis.   To do this properly one would have to Kramers-Kronig transform $|R_{xx} \pm i R_{xy}|$, whereas in usual FTIR only $|R_{xx}|$ and $| R_{xy}|$ are measurable.}.  Such experiments in high magnetic field have only been possible recently~\cite{baydin2021time}. 

Our system combines pulsed magnetic fields with time-domain THz spectroscopy (TDTS) using an electronically-controlled optical sampling (ECOPS) system (as described in Ref.~\cite{post2021observation}), wherein the timing delay between the two ultrafast optical pulse trains that drive the THz emitter and receiver can be electronically modulated quickly. When synchronized to the magnetic field, a complete TDTS spectrum can be sampled in 30 $\mu$s with approximately 7 spectra taken in the $\approx$ 5 ms long pulse duration.  To further improve the signal-to-noise, at each temperature we averaged THz data from multiple magnet pulses (typically 10-20 pulses, which required 2-4 hours per temperature).  The magnet has a 15 mm bore, comprising 144 windings of high-strength CuAg wire, and is powered by a purpose-built 20 kJ capacitor bank. The field profile of a typical 31 T pulse is shown in Ref.~\cite{post2021observation}.  The magnetic field is applied along the \textit{c}-axis of the sample, in Faraday geometry in which the magnetic field is parallel to the THz propagation vector.

We performed a first set of measurements in zero magnetic field using a different spectrometer in order to get the intrinsic complex transmission of the sample without the contribution of the substrate. The transmitted time-domain THz electric fields are Fourier transformed and then ratioed to the Fourier transform of the fields through a blank substrate to get $T(\omega,0)$. We then measured the thin films in the high field spectrometer, including a zero-field measurement for referencing (there is no contribution of the substrate to the transmitted signal in magnetic field), yielding $T(\omega,B)$.

Transmission through the sample is governed by a $2\times2$ Jones matrix.  For a sample with tetragonal symmetry the transmission matrix's components are constrained such that $T_{xx}= T_{yy}$ and $T_{xy}= -T_{yx}$~\cite{armitage2014constraints}.  With the polarized E field in the $x$ direction, access to $T_{xy}$ in the current experiment is achieved by
incorporating two linear polarizers (P1 and P2) into the THz beam path.   See Ref.~\cite{post2021observation} for an experimental schematic. When P2 is rotated $\pm 45^\circ$ with respect to P1, then $T( \pm 45^\circ)  = (T_{xx} \pm T_{xy} )/ \sqrt{2}$, from which we can extract $T_{xx}$ and $T_{xy}$.   Then the effective transmissions in the right ($r$) and left ($l$) circular bases could be derived from the relation $T_{rr/ll} = T_{xx} \pm i T_{xy}$ ~\cite{post2021observation}.
The circularly polarized complex transmission coefficient (and associated conductivity) at fields up to 31 T were obtained in the frequency range from 0.4 to 1.4 THz. Such experiments with a continuum of frequencies give more information than previous IR Hall experiments that only measured a few discrete frequencies in the infrared regime~\cite{schmadel2007infrared,jenkins2010terahertz}.  As $r$ and $l$ are the eigenbases for transmission through a tetragonal material in Faraday geometry, one can directly invert the complex transmission in the circular basis to obtain the complex conductivity in the same basis via the standard transmission formula for a thin film on a substrate:  

\begin{equation}
\tilde{T}(\omega)_{rr/ll}=\frac{1+n}{1+n+Z_0\sigma_{rr/ll}( \omega)t} e^{i\Phi_s}.
\end{equation}
Here $\Phi_s$ is the phase accumulated from the small difference in thickness between the sample and reference substrates, $t$ is the film thickness, $Z_0$ is the impedance of free space ($\approx$377 Ohms), and $n$ is the substrate index of refraction.  Because one measures a complex transmission function, the inversion to complex conductivity is done directly without Kramers-Kronig transformation.  Note again that the direct optical measurement of $\sigma_{rr/ll}$ in magnetic field is only possible with a phase sensitive experiment like TDTS.

\begin{figure*}[t]
\centering
\includegraphics[width=1.0\textwidth]{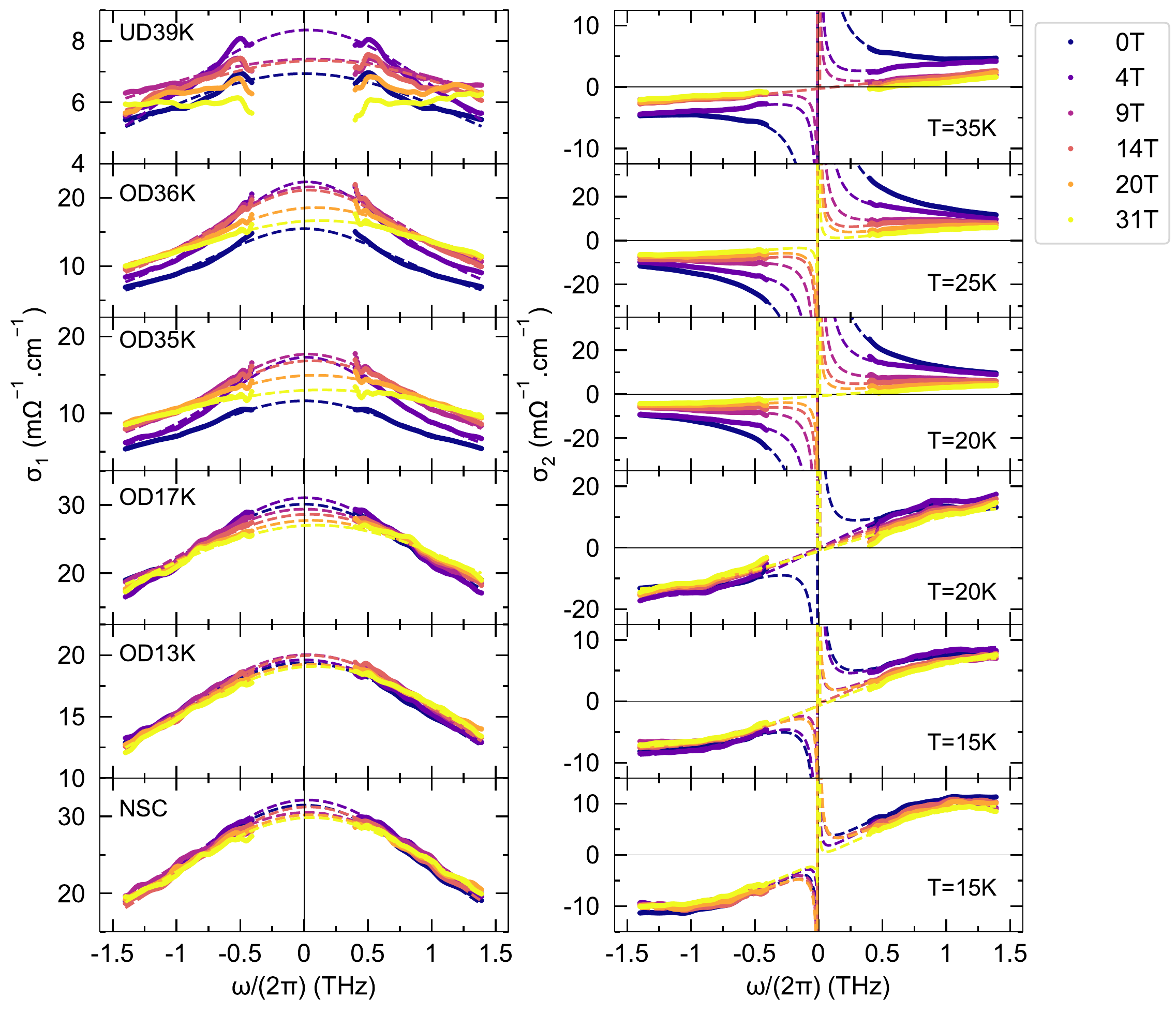}
\caption{Circularly-polarized complex conductivity of La$_{2-x}$Sr$_x$CuO$_4$ thin films as a function of frequency (negative and positive frequencies correspond to $\sigma_{ll}$($\omega$) and $\sigma_{rr}$($\omega$) respectively) at various magnetic fields up to 31T. Six different dopings, covering the whole doping range,
are represented (as labelled in Table. \ref{samples}), at one temperature near their respective $T_c$. Left and right panels correspond to the real and imaginary parts of the conductivity respectively.  Dashed lines show fits to the complex conductivity using a cyclotron-active two-fluid model of Eq.\ref{CircCond}. Note that depending on the measurement, the value of the mean magnetic field represented by one specific color can vary a little (due to shot-to-shot variation in the pulse fields), such that there is an error bar of about $\pm$~1T for each magnetic field value in the legend.} 
\label{Fig.Conductivity}
\end{figure*}

The CR phenomenon itself is also best described in the circular basis for conductivity, where $rr$ and $ll$ conductivities can be parametrized as the response to positive and negative frequencies.   As such we define the quantity $\sigma_{rr/ll}$ as a single continuous response function defined for both positive and negative frequencies.  $\sigma_{rr/ll} $ is related to the usual longitudinal and transverse conductivities through the relations $\sigma_{xx}(\omega)  = [\sigma_{rr/ll}(\omega) + \sigma_{rr/ll}^*(-\omega)]/2 $ and $\sigma_{xy} (\omega) = [\sigma_{rr/ll}(\omega) - \sigma_{rr/ll}^*(-\omega)]/2i $, where $\sigma_{xx}$ and $\sigma_{xy}$ are defined only for positive frequencies and $^*$ indicates complex conjugation.  Within the Drude model the $\sigma_{rr/ll}$ conductivity is
\begin{equation}
\sigma_{rr/ll}(\omega) = i\epsilon_0
\left( \frac{\omega^2_{\rm p,n}}{\omega - \omega_{\rm c} +i \Gamma } + \frac{\omega_{\rm p,s}^2}{\omega} - \omega (\epsilon_{\infty}-1) \right),
\label{CircCond}
\end{equation}
where $\sigma_{rr/ll}$ is the complex conductivity for right- or left-handed circularly polarized light, $\omega^2_{\rm p,n/s}$ is the squared plasma frequency for normal or superconducting carriers that is equal to $ \frac{n e^2}{m \epsilon_0}$, $\omega_c$ is the cyclotron frequency, $\Gamma$ is the scattering rate, and $\epsilon_{\infty}$ is the background dielectric constant.  In this expression, $rr$ and $ll$ conductivities correspond to arguments with positive and negative frequencies respectively.  The first term describes the transport contribution of the normal carriers (e.g. the Drude model).  Note that in the $r$ and $l$ basis this function has the simple functional form of a Lorentzian peak shifted in positive or negative frequency directions by the cyclotron frequency.  This is a principal reasons that the $r$ and $l$ bases are superior over the $x$ and $y$ basis for analysis and display of the data.   The functional form in the $x$-$y$ basis is much more complicated.  The second term is the contribution of any superconducting carriers.   Although our primary interest is the normal state, we include it in the fits to account for the small fluctuating superconducting signal that can exist right above $T_c$ and at small fields.  This two-fluid model has been extensively used to fit and describe the THz spectra of the cuprates~\cite{bilbro_possibility_2011,bilbro_temporal_2011,mahmood_locating_2019}.  However, for the temperatures at which the measurements were conducted, our maximum field of 31 T is far larger than $H_{vs}$ (the vortex-solid melting field) in all samples, so the contribution of the superconducting term to conductivity is negligible above 5T.  Note that in these measurements, we did not measure the same temperatures on all dopings --  the experiments are time intensive -- and instead concentrated on the lowest temperatures such that the scattering rate was small enough to observe a CR, but that we could still suppress any superconductivity at low fields.  The temperature dependencies to the masses appears small.  Note the use of a frequency-independent scattering rate $\Gamma$ is justified at our low frequencies, as any frequency dependence over our small spectral range is much smaller than the scattering rate’s overall scale.  The third term is the background dielectric contribution, which arises from excitations well out of our measurement range.  Our results are largely insensitive to the value of $\epsilon_{\infty}$.

\section{III. Results}

In Fig. \ref{Fig.Conductivity} we plot the measured complex conductivity in the circular basis as a function of positive and negative frequencies $\sigma_{rr/ll} (\omega)$ for various samples and magnetic fields.  Negative and positive frequencies correspond to $\sigma_{ll}$($\omega$) and $\sigma_{rr}$($\omega$) respectively.  Only one temperature (at a temperature near $T_c$) is shown for each sample.  The different colors correspond to different fields. The dotted lines are the fits to the model described above (all parameters with the exception of $\epsilon_{\infty}$ are free). The large upturns at low frequency in $\sigma_2$ at small fields indicate the presence of superconducting fluctuations. At small applied fields this upturn is eliminated, indicating the superconducting term is suppressed. Similarly, the field dependence of the normal state spectral weight ($\propto$~$\omega_{p}^2 = \frac{n e^2}{m \epsilon_0}$ in the Drude model, given by the area of the conductivity curves) plateaus above 4-9 T~\cite{post2021observation} (this is why in some samples at small field the effect that we observe is an increase in $\sigma_1$ until we reach the normal state).  With increasing field, we observe a small, but systematic shift in $\sigma_1(\omega)$ to positive frequencies, along with a slight broadening of the Drude peak for most dopings.  The width of the Drude peak corresponds to the scattering rate of the charge carriers. The broadening of the peak therefore implies an increase in scattering as the field is applied (this effect is mostly observed in the smallest dopings).  The field-driven asymmetry is due to the cyclotron shift of the metallic Drude peak.  Note that even at the largest fields the shifts are small and only resolvable due to being able to measure the $rr$ and $ll$ conductivities separately. The shifts would not be resolvable in the real part of $\sigma_{xx}$ by itself, as it is the average of $\sigma_{rr}$ and $\sigma_{ll}$.
 
\begin{figure}[h!]
\centering
\includegraphics[width=0.95\columnwidth]{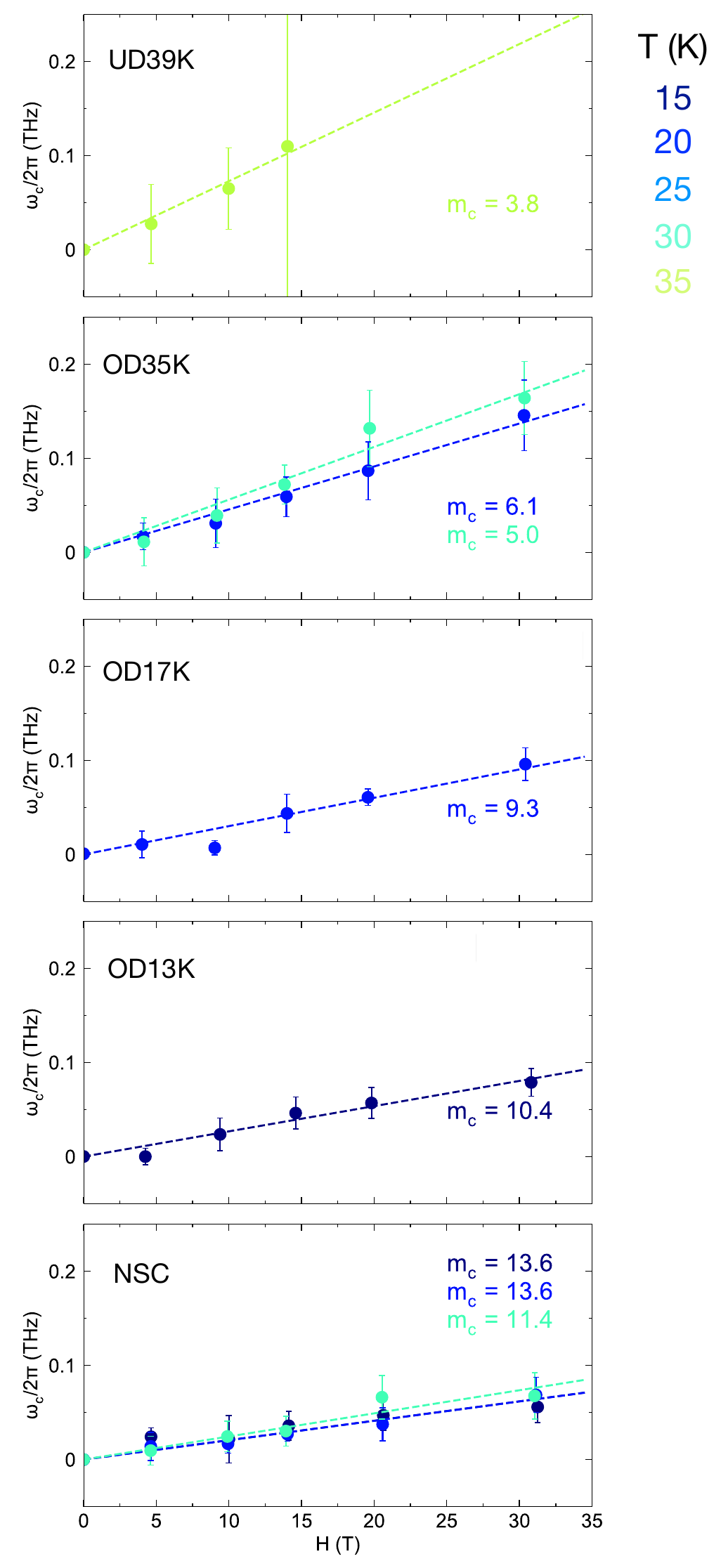}
\caption{Cyclotron frequency as a function of magnetic field for five different samples with dopings covering the range $p = 0.13-0.26$. The different temperatures are indicated using different colors. The extracted cyclotron masses are given in units of the bare electron mass. We could not obtain the cyclotron frequency at fields greater than 14T for the underdoped sample due to the inability to fit the conductivity data $\sigma(\omega)$ with our model at these fields and frequency range.}
\label{Fig.Cyclotron_frequency}
\end{figure}

In Fig. \ref{Fig.Cyclotron_frequency} we plot the values of $\omega_c$ extracted from fits to the conductivity as a function of magnetic field (some samples with multiple temperatures).  All displayed data come from samples near the transition to the normal state and had whatever residual (possibly fluctuating) superconductivity suppressed at small fields.  A positive shift of the cyclotron frequency, corresponding to hole-type carriers, was observed in all studied samples spanning from the slightly underdoped to highly overdoped ($T_c$~=~5K).  Moreover, there was no obvious deviation from linearity in $\omega_c(B)$ (as expected for the field-dependence of cyclotron frequency $\omega_c$ = $eB/m_c$), meaning that we detect no field-dependent Fermi surface reconstructions for $H\leqslant$31T.  We believe that the different form of the conductivity spectra observed above 15T in LSCO UD39K may be due to the presence of quasi-static magnetism (slow freezing of spins) that could add to the scattering as discussed in Ref.~\cite{frachet_hidden_2020} and leading to a very flat real part of conductivity as a function of frequency.   In this regard, it is important to measure even more underdoped samples, but their larger scattering rates will require magnetic fields in excess of the currently achievable 31T.

\begin{figure}[h!]
  \begin{center}
    \includegraphics[width=1.05\columnwidth]{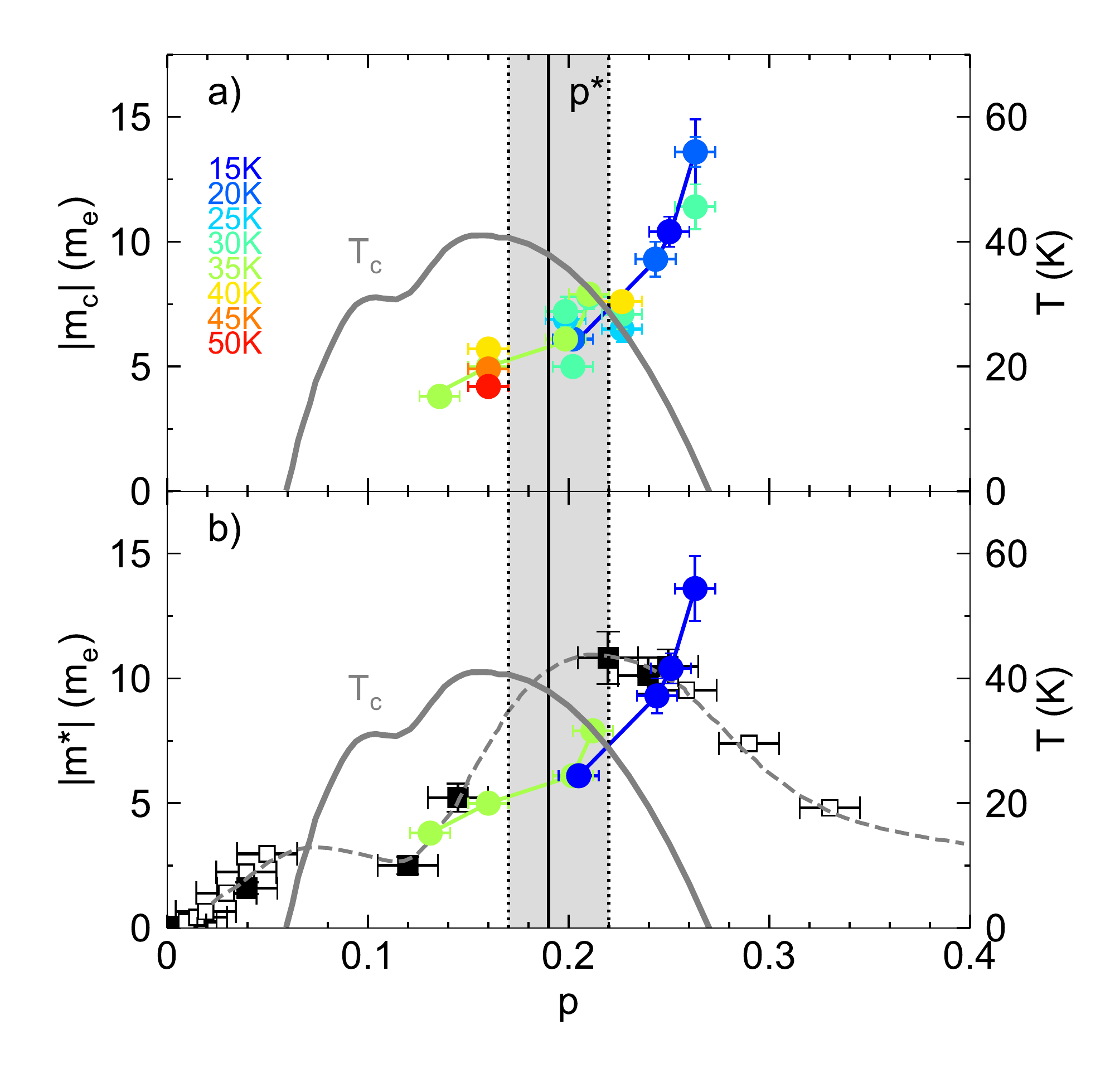}
	\caption{\textit{Top}: Doping dependence of the cyclotron mass (in units of the bare electron mass) at the indicated temperatures.  The gray solid line shows the $T_c$ value (that can be read on the right $y$-axis) as a function of doping for thin films.
	\textit{Bottom}: Black open and closed squares are effective masses estimated from the electronic specific heat on crystals in the normal state compiled from different sources as shown in Ref.\onlinecite{girod_normal_2021}.  The dashed gray line is a guide to the eye.	Circles represent the cyclotron mass at the lowest measured temperature for each sample (using the same color code as top figure).  The vertical solid line along with the vertical dotted lines represent (with error bars) the position of the pseudogap critical point.  Horizontal error bars of the specific heat data represent the uncertainty in doping levels when comparing films and crystals Fermi surface volumes in ARPES~\cite{Zhong2022}.}
	\label{Fig.Cyclotron_mass}
  \end{center}
\end{figure}

\begin{figure}[h]
\centering
\includegraphics[width=0.95\columnwidth]{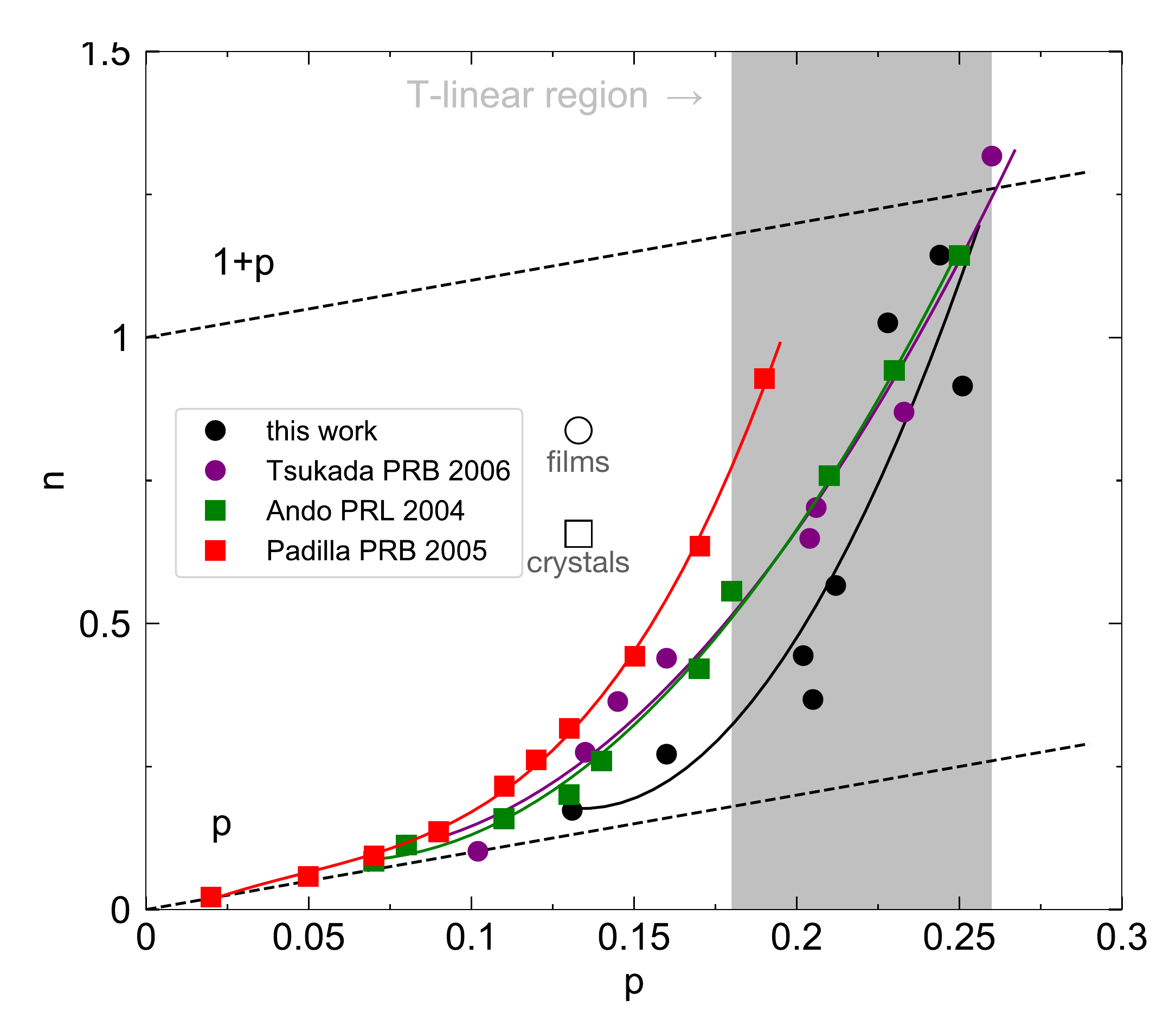}
\caption{Inferred filling from a number of Hall measurements at low temperature~\cite{ando_evolution_2004,padilla2005constant,tsukada_negative_2006} as compared to the densities extracted from the present measurements (the values are taken at the smallest temperature for each sample).
One can see the crossover from a dependence that goes from the linear $n \sim p$ at low doping to a faster dependence at higher doping that tends to $n \sim 1 + p$. We do not show dopings above $p>0.25$ as the values diverge due to the sign change of $R_H$. The shaded region corresponds to the region in doping where $T$-linear resistivity is observed as $T\rightarrow$0 \cite{cooper_anomalous_2009}. } 
\label{sub:nH}
\end{figure}

In Fig. \ref{Fig.Cyclotron_mass}a, we convert the cyclotron frequency into a cyclotron mass using the standard relation $\omega_c$ = $eB/m_c$ and plot it as a function of the estimated doping (for various temperatures, each color representing a specific temperature). The doping is estimated using $T_c$ of the sample and the empirical law of Ref.~\cite{presland_general_1991} as discussed above. Again, note that we did not measure the same temperatures on all dopings, as we aimed for the lowest temperatures such that the scattering rate was small enough to observe a CR, but that we could still suppress any superconductivity at low fields.  This introduces insignificant uncertainty as the temperature dependence -- if any -- is small.  One can observe a clear trend of a mass that monotonically increases from 3.8 $m_e$ to 13.6 $m_e$ over the range from slightly underdoped to very overdoped.  Surprisingly, we observed no sign change of the cyclotron mass despite the doping-induced Lifshitz transition at $x=0.205$ where the Fermi surface changes from a hole pocket at ($\pi,\pi$)  to an electron pocket at ($0,0$)~\cite{yoshida_systematic_2006,Zhong2022}.  As discussed further below this is surprising in conventional theories as the sign of the cyclotron resonance is a topological quantity sensitive to whether the Fermi surface closes around holes or electrons.

Specific heat measurements in high magnetic fields revealed a peak in the effective mass close to $p^*$~\cite{horio_three-dimensional_2018,girod_normal_2021},  the interpretation of which is complicated by the close proximity to both the Lifshitz transition at $p=0.205$ and $p^*$ at 0.19.  Through fits to the ARPES dispersion,  Refs.~\cite{horio_three-dimensional_2018,girod_normal_2021} claim that this peak is too large to come only from the enhanced density of states at E$_F$ near the Lifshitz transition and that critical fluctuations at the $p^*$ quantum critical point contribute to the mass enhancement.  In contrast,  recently  Ref.~\cite{Zhong2022} has claimed that a new parametrization of the LSCO ARPES dispersion incorporating a doping-dependent band structure {\it can} reproduce the peak in the heat capacity and the heat capacity anomaly can be solely attributed to the Lifshitz transition.  Although different effective masses can be defined even within Fermi liquid theory, the expectation is that the thermodynamic mass measured in heat capacity is the same as the quasiparticle mass measured with ARPES~\cite{pines1966theory}.  To compare the masses measured with heat capacity and cyclotron resonance, we converted the measured electronic specific heat coefficient in crystals (from Ref.~\cite{girod_normal_2021}) into an effective mass $m^*$, (using the relation between $\gamma$ and $m$ for a 2D metal with a single Fermi surface $\gamma$~=~($\pi N_A k_B^2$/3$\hbar^2$)$a^2 m^*$, where $a$ is the lattice parameter of LSCO) and plotted it alongside cyclotron mass in Fig. \ref{Fig.Cyclotron_mass}b.   One can see that the cyclotron mass is systematically smaller than the thermodynamic mass for dopings $p<0.25$, but it is greater at higher dopings.

In Fig.~\ref{sub:nH} we show the measured charge
carrier density per Cu ion determined from the fitted spectral weights and measured mass.  We compare these values with the values of $n_H$ from transport studies, in both thin films and single crystals, and observe a similar trend in doping (note that the work from Balakirev \textit{et al.} showing a small anomaly in the monotonous dependence at very low temperatures \cite{balakirev_quantum_2009} is not used here). One can see a general increase of $n$ with doping from about $p$ to $1+p$, as previously observed in several hole-doped cuprates \cite{badoux_critical_2016,collignon_fermi-surface_2017}, but instead of reaching $1+p$ at $p\approx p^*$, it seems here that this value is reached at higher doping, closer to the end of the $T$-linear resistivity region and the superconducting dome. This singularity in the doping dependence of $n$ is reminiscent of the singularity in the appearance of $T$-linear resistivity as $T\rightarrow 0$ in LSCO (occurring up to dopings largely above $p^*$ instead of just $p\approx p^*$ \cite{cooper_anomalous_2009}).

\section{IV. Discussion}

Our data present a number of observations at odds with the simple expectation.  We find the sign of the cyclotron mass is unchanged throughout our entire doping range despite indications from ARPES of a hole- to electron-like Lifshitz transition~\cite{yoshida_systematic_2006,Zhong2022} at $x \approx 0.205$.  Moreover, we find $m_c$ is systematically less than the thermodynamic mass for samples near the $p^*$ region (from optimally doped up to $p\approx0.25$) and exceeds it for $p>0.25$, with no sign of decrease up to dopings near the end of the superconducting dome.  As discussed above, this disagreement can be contrasted with a recent ARPES study~\cite{Zhong2022} that has shown that a doping dependent tight-binding parametrization of the band structure, (excepting the most underdoped samples) has a calculated heat capacity that matches the experimental one even through the Lifshitz transition.  We see no sign of a divergence of the mass near $p^*$ nor near the Lifshitz transition, showing that any singularity -- if it exists -- is not strong enough to affect the cyclotron mass.

The enhancement of fermionic masses is a well-known consequence of strong correlations. However the degree of enhancement can depend on the particular experiment.  The amplitude of quantum oscillations (Shubnikov–de Haas and de Haas–van Alphen) in high magnetic fields and the fermionic contribution to the heat capacity measure a thermodynamic mass that reflects the density of states~\cite{pines1966theory,shekhter2017thermodynamic}.  Angle-resolved photoemission measures a mass corresponding to the renormalized quasiparticle dispersion.   Within Fermi liquid theory with momentum-independent scattering these are the same~\cite{merino2000cyclotron}.  This mass incorporates the effects of both electron-electron and electron-lattice interactions.  In contrast, within Fermi liquid theory, measurements of the susceptibility or compressibility are sensitive to a density of states that reflects only electron-electron interactions but not electron-phonon interactions~\cite{pines1966theory}\footnote{There may be some evidence for this experimentally.  Ref.~\cite{girod_high_2020} points out that Ref.~\cite{kawasaki_carrier-concentration_2010} indicates a different density of states near $E_F$ in Bi$_2$Sr$_{2-x}$La$_x$CuO$_{6+\delta}$ as measured in the spin susceptibility via the Knight shift than the heat capacity indicates.}. As mentioned above, Kohn's  proof~\cite{kohn1961cyclotron} showed that in a Galilean invariant system electron-electron interactions do not modify the cyclotron mass.  These considerations do not apply to materials like cuprates at conventional charge densities, where disorder, non-parabolicity, and umklapp scattering breaks Galilean invariance strongly~\cite{macdonald1989cyclotron,kanki1997theory}.   However Kohn's result establishes that the cyclotron mass can be sensitive to interactions in a different fashion than other probes and therefore comparing masses may bring new insight into the dominant interactions.

Our observation that the cyclotron frequency is positive for all our measured dopings is reminiscent of the dc Hall constant ($R_H$) that does not change sign until $x>0.3$ \cite{tsukada_negative_2006}.  Within a Boltzmann transport approach $R_H$~\cite{ong1991geometric} is determined by the anisotropic Fermi surface contributions weighted by the mean free path $l_{\bf{k}} = v_{\bf{k}}  \tau_{\bf{k} } $ of the convex and concave portions of the Fermi surface.  In diagrammatic approaches, vertex corrections are important when calculating $R_H$, and the relative portion of the Fermi surface inside and outside the magnetic Brillouin zone contribute differently~\cite{kontani1999hall}.  However, our result may be considered more surprising as -- unlike $R_H$ which can depend on scattering -- the sign of the cyclotron mass (as inferred from a finite frequency resonance) in the conventional treatment is a {\it topological} quantity of a closed Fermi surface, measuring whether the Fermi surface encloses electrons or holes ($m_c = \frac{\hbar^2}{2 \pi } \frac{\partial A}{\partial E} |_{E_F} $).  The disagreement between the measured thermodynamic and cyclotron mass may indicate either strongly anisotropic scattering or explicitly non-Fermi liquid-like physics.    
Although the effect of anisotropic scattering on the sign and magnitude of the Hall effect has been discussed extensively~\cite{ong1991geometric,kontani1999hall,tsukada_negative_2006}, we know of only a single study~\cite{kanki1997theory} that explicitly calculated the cyclotron frequency in the presence of momentum-dependent scattering.\footnote{Here it is important to make a distinction between calculations of the frequency of resonance of cyclotron motion with ones that calculate the coefficient of the argument in the Lifshitz-Kosevich treatment of quantum oscillations.   The latter ``cyclotron frequency", although algebraically the same, can have a different effective mass.}  Ref.~\cite{kanki1997theory} calculated the vertex corrections within a diagrammatic treatment of the Hubbard model and showed umklapp scattering tended to enhance the cyclotron mass above the band value.   In extreme situations where umklapp scattering is particularly enhanced (e.g. near half-filling), the cyclotron mass can even exceed the thermodynamic one.  Although important, this study used -- among other aspects -- only the simplest nearest-neighbor tight-binding model dispersion.  Going forward, it is important to perform such calculations using more physical band structure parameters.  It is also important to model the phenomenon of cyclotron resonance in the presence of anisotropic scattering more thoroughly.

We can compare our results with both recent and older measurements of the effective mass from optical spectral weight (which is proportional to the squared plasma frequency $ \frac{n e^2}{m \epsilon_0})$ analysis.  At zero magnetic field, optical conductivity can define a renormalized mass as a ratio of the total intraband spectral weight to the total spectral weight in the low frequency coherent Drude-like peak.  This gives the ratio of the renormalized mass to the band mass $(m_b)$.  Alternatively if a measure or assumption can be made for the charge density $n$ then the mass can be derived from the spectral weight of the low frequency coherent Drude-like peak itself.  Using a combination of low temperature Hall effect data and low temperature optical conductivity in LSCO  Ref. \onlinecite{padilla2005constant} 
found that up to dopings of 0.17 the mass ratio $m^*/m_e$ was approximately constant at 5.   This observation is at odds with the heat capacity experiments that (in a band model) infer the mass increases with carrier density~\cite{girod_high_2020}.  More recently, Ref. \onlinecite{michon2021spectral} characterized the spectral weight in the low frequency coherent Drude-like peak in the optical conductivity and noted that if one {\it assumed} $x\sim p$ for underdoped samples and $x = 1+p$ for overdoped samples, then the optical masses roughly follow those extracted from the heat capacity.   However it is important to note that this result is in strong disagreement with our result of a (cyclotron) mass that increases monotonically towards the overdoped superconducting phase boundary.



\section{V. Conclusion} 

Over a wide doping range ($p=0.13-0.26$) of LSCO, we have measured the frequency dependence of the complex conductivity in the right- and left-handed circular polarization channels using time-domain THz spectroscopy coupled to a pulsed magnet. Data at magnetic fields up to 31T and different temperatures revealed the cyclotron resonance of the charge carriers.  A fit of the complex conductivity to the Drude expressions for magneto-conductivity allowed the extraction of crucial information such as the cyclotron mass, scattering rate and density of carriers.  There are a number of notable aspects to the data.  The sign of the cyclotron mass is unchanged throughout the range of dopings from $p=0.13$ to 0.26, despite indications from ARPES of a hole- to electron-like Lifshitz transition at  $x=0.205$~\cite{yoshida_systematic_2006,Zhong2022}.  We also find the cyclotron mass is systematically less than the thermodynamic mass for samples with $p<0.25$ and exceeds it for $p>0.25$. In the latter regard it is important to note that the mass appears to be enhanced on the approach to the overdoped metal-superconductor quantum phase transition.   It could be the case that this is a signature of quantum criticality.

These results are striking because for weak scattering within conventional theoretical treatments the sign of the finite frequency cyclotron resonance for a 2D metal is (unlike the sign of the Hall effect) a topological quantity that reflects the global curvature of the Fermi surface.   In a Fermi liquid with momentum-independent scattering, the cyclotron mass should reflect the density of states near-$E_F$ as measured in heat capacity~\cite{merino2000cyclotron}.   However the phenomenon of cyclotron resonance is under investigated in the regime of strong anisotropic scattering.   Most theoretical treatments were developed to explain results on low density semiconductor systems with weak momentum dependent scattering~\cite{macdonald1989cyclotron}.   It is therefore worthwhile to further investigate cyclotron resonance in the presence of anisotropic scattering and lattice effects.  The technique (in the microwave range) has been applied to heavy-fermions and ruthenates~\cite{tonegawa2012cyclotron,hill2000cyclotron}, but there the physics is arguably more complicated due to multi-band effects.  We propose that cuprates are a better platform for understanding the interplay between correlations and lattice effects in cyclotron resonance as the materials are single-band and the Fermi surfaces are  simpler.   With regards to non-Fermi liquids there have been very few calculations of cyclotron resonance in strongly interacting systems, but there are systems like composite fermions (2DEG in a large magnetic field) where the generated cyclotron mass is completely unrelated to either the free electron mass or the band mass.  Charge carriers move like electrons along closed cyclotron orbits, but with a mass generated solely from electron–electron interactions~\cite{kukushkin2002cyclotron}.  It would also be important to address how large scale inhomogeneity would impact the measurement of cyclotron resonance.   Such inhomogeneities can influence the global longitudinal dc magnetoresistance by mixing the local and transverse responses ~\cite{parish2005classical,singleton2020temperature} and they should be investigated in the present context.

For future experiments it will be important to increase the available field range to 50-60 Tesla to access this physics in more underdoped samples with larger scattering rates, and look for possible Fermi surface reconstructions.  Higher fields will also allow superconductivity to be suppressed at lower temperatures and hence a wider range of normal state behavior can be investigated.  Among other aspects it will be interesting to look for the ln$(T_0/T)$ mass renormalizations seen in the heat capacity~\cite{girod_high_2020}, which are believed to be a sign of quantum criticality.  It would also be extremely interesting to perform these experiments in high quality films of electron-doped cuprates, where interesting Hall effect anomalies have been found, and quantum oscillations have been measured~\cite{armitage2010progress,helm2009evolution,tagay2021bcs}

Finally we point out that these experiments are emblematic of advances enabled by THz spectroscopy coupled to high pulsed magnetic fields.  In addition to problems like this in correlated superconductivity, there are vast applications of these techniques to magnetic materials and topological systems~\cite{baydin2021time}.

\section*{Acknowledgments}

 AL and NPA were supported by the Quantum Materials program at the Canadian Institute for Advanced Research, the NSF DMR 1905519, and the Gordon and Betty Moore Foundation’s EPiQS Initiative through Grant No. GBMF-9454. Work at the National High Magnetic Field Laboratory was supported by National Science Foundation (NSF) DMR-1644779, the State of Florida, and the U.S. Department of Energy (DOE).  SAC acknowledges support from the US DOE Office of Basic Energy Sciences `Science of 100T' program.  Work at Brookhaven National Laboratory was supported by the DOE, Basic Energy Sciences, Materials Sciences and Engineering Division. X. H. is supported by the Gordon and Betty Moore Foundation’s EPiQS Initiative through grant GBMF9074. We gratefully acknowledge G. Granroth for providing the CuAg wire used in the pulsed magnet.   We would also like to thank D. Basov, S. Chakravarty, G. Grissonnanche, S. Kivelson,  R. Laughlin, I. Martin, R.D. McDonald, B. Ramshaw, T. Senthil, J. Singleton, A. Shekter, and D. van der Marel for insightful conversations, and Z.-X. Shen and Y. Zhong for discussions and getting an early look at their -- at the time -- unpublished data~\cite{Zhong2022}.

\bibliography{library} 		
\bibliographystyle{apsrev4-1} 		


\end{document}